\documentclass[english]{article}
\usepackage[T1]{fontenc}
\usepackage[latin9]{inputenc}
\usepackage{amstext}
\usepackage{amssymb}
\usepackage{graphicx}

\makeatletter
\usepackage[noend]{algpseudocode}
\usepackage{algorithm}
\usepackage{bbm} %für identity
\algnewcommand{\algorithmicgoto}{\textbf{go to}}%algorithms
\algnewcommand{\Goto}[1]{\algorithmicgoto~\ref{#1}}%

\usepackage[final,nonatbib]{neurips_2020_ml4ps}

\usepackage{url}% simple URL typesetting
\usepackage{amsfonts}% blackboard math symbols
\usepackage{nicefrac}% compact symbols for 1/2, etc.
\title{Temperature-steerable flows}

\author{%
  Manuel Dibak\footnotemark[1]\\
  FU Berlin\\
  Berlin, Germany\\
  \texttt{manuel.dibak@fu-berlin.de}
\And
  Leon Klein\thanks{both authors contributed equally}\\
  FU Berlin\\
  Berlin, Germany\\
  \texttt{leon.klein@fu-berlin.de}
\And
  Frank Noé\\
  FU Berlin\\
  Berlin, Germany\\
  Rice University\\
  Houston, Texas\\
  \texttt{frank.noe@fu-berlin.de}
}

\makeatother

\usepackage{babel}
\begin{document}
\maketitle
\begin{abstract}
Boltzmann generators approach the sampling problem in many-body physics
by combining a normalizing flow and a statistical reweighting method
to generate samples of a physical system's equilibrium density. The
equilibrium distribution is usually defined by an energy function
and a thermodynamic state, such as a given temperature. Here we propose
temperature-steerable flows (TSF) which are able to generate a family
of probability densities parametrized by a choosable temperature parameter.
TSFs can be embedded in a generalized ensemble sampling framework
such as parallel tempering in order to sample a physical system across
thermodynamic states, such as multiple temperatures.
\end{abstract}

\section{Introduction}

Sampling equilibrium states of many-body systems such as molecules,
materials or spin models is one of the grand challenges of statistical
physics. Equilibrium densities of system states $x$ are often given
in the form
\begin{equation}
p_{X}(x)\propto\exp(-u(x)),\label{eq:equilibrium_density}
\end{equation}
where $u(x)$ is a reduced (unit-less) energy that combines the system's
potential $U(x)$ (if momenta are of interest we have the Hamiltonian
energy instead) with thermodynamic variables that define the statistical
ensemble. In the canonical ensemble the reduced energy is given by
$u(x)=U(x)/\tau$ where the thermal energy $\tau=k_{B}T$ is proportional
to the temperature $T$ and $k_{B}$ is the Boltzmann constant. In
order to reach all thermodynamic states, we would like to sample a
family of densities parameterized by the thermodynamic control variables
-- in the canonical ensemble:
\begin{equation}
p_{X}(x;\tau)\propto\exp(-U(x)/\tau).\label{eq:canonical_density}
\end{equation}
The most common approach to sample densities (\ref{eq:equilibrium_density})
in physics and chemistry are Markov Chain Monte Carlo (MCMC) or Molecular
Dynamics (MD) -- both proceed in steps, making small changes to $x$
at a time, and guarantee that the target density (\ref{eq:equilibrium_density})
will be sampled asymptotically but possibly after a very long or intractable
simulation time. Generalized ensemble methods such as parallel tempering
(PT) simulate multiple copies of the system at different temperatures
(or other thermodynamic states) and perform Monte-Carlo exchanges
between them \cite{swendsen1986replica,Geyer1991MarkovCM,Hukushima1996ExchangeMC}.
These approaches can converge faster than direct MCMC/MD, and sample
the generalized ensemble (\ref{eq:canonical_density}) directly.

Recently, there has been a lot of interest to train normalizing flows
\cite{TabakVandenEijnden_CMS10_DensityEstimation,RezendeEtAl_NormalizingFlows,DinhDruegerBengio_NICE2015,dinh2016density,Papamakarios2019NormalizingFF,kobyzev2020normalizing}
to sample densities of many-body physics systems such as (\ref{eq:equilibrium_density})
directly without having to run long, correlated simulation chains.
Normalizing flows transform an easy to sample prior distribution $p_{Z}(z)$,
e.g. a multivariate normal distribution, via a transformation $x=f(z)$
to the output distribution $p_{X}(x)$. If $f(z)$ is invertible,
$p_{X}(x)$ can be computed by the change of variable formula
\begin{equation}
p_{X}(x)=p_{Z}(z)\left|\det J_{f}(z)\right|^{-1},\label{eq:change_of_variable}
\end{equation}
where $\left|\det J_{f}(z)\right|^{-1}$ is the inverse of the Jacobian.
Boltzmann Generators (BGs) \cite{noe2019boltzmann} combine normalizing
flows to minimize the distance between (\ref{eq:equilibrium_density})
and (\ref{eq:change_of_variable}) with a statistical reweighting
or resampling method to generate unbiased samples from (\ref{eq:equilibrium_density}).
This and similar approaches have been used to sample configurations
of molecular and condensed matter systems \cite{noe2019boltzmann,wu2020stochastic},
spin models \cite{Li2018NeuralNR,Nicoli2020AsymptoticallyUE} and
gauge configuration in lattice quantum chromodynamics \cite{Albergo2019FlowbasedGM,Boyda2020SamplingUS}.

In order to address the problem of sampling generalized ensembles
such as (\ref{eq:canonical_density}), we here take a first stab at
developing flow architectures that can be steered by thermodynamic
variables. Steerable flows are related to equivariant flows which
maintain group transformations such as rotation and permutation throughout
the flow \cite{Khler2020EquivariantFE,Rezende2019EquivariantHF,zhang2018monge}.
Specifically, we develop temperature-steerable flows (TSF) that correctly
parametrize the distribution $p_{X}$ by a temperature variable. We
evaluate our model on a test system and Alanine Dipeptide, show that
the TSF significantly improves the temperature scaling, is capable
of producing samples close to equilibrium at different temperatures,
and can be turned into a multi-temperature BG by using it as a proposal
density in a PT framework.

\section{Temperature-steerable flows}

\paragraph{Temperature scaling}

Up to a normalization constant, a change to temperature $\tau'$ of
the Boltzmann distribution corresponds to raising it by the power
of $\kappa=\tau/\tau'$, $p_{X}^{\tau'}(x)\propto\left[p_{X}^{\tau}(x)\right]^{\kappa}.$
Using Eq. (\ref{eq:change_of_variable}) we observe that the temperature
scaling is exact, if for any two temperatures $\tau,\tau'$
\begin{equation}
p_{Z}^{\tau'}(z)\left|\det J_{f_{\tau'}}(z)\right|^{-1}\propto\left[p_{Z}^{\tau}(z)\left|\det J_{f_{\tau}}(z)\right|^{-1}\right]^{\kappa}.\label{eq:scaling_condition}
\end{equation}
The proportionality in the prior distribution can be matched by selecting
a Gaussian prior $p_{Z}^{\tau}(z)=\mathcal{N}(z|\vec{0},\tau)$, which
fulfills $p_{Z}^{\tau'}(z)\propto\left[p_{Z}^{\tau}(z)\right]^{\kappa}$.
This results in a condition on the Jacobian of the flow $\left|\det J_{f_{\tau}}(z)\right|^{\kappa}\propto\left|\det J_{f_{\tau'}}(z)\right|$.
As a consequence, volume-preserving flow layers, i.e. $\left|\det J_{f_{\tau'}}(z)\right|=1$,
such as NICE \cite{DinhDruegerBengio_NICE2015} fulfill this condition
trivially. However, the condition is also fulfilled by constant Jacobians,
i.e. $\left|\det J_{f_{\tau'}}(z)\right|=\text{const.}$ which we
make use of to develop more expressive TSFs.

\paragraph{An architecture for temperature-steerable flows}

\begin{figure}
\centering{}\includegraphics[width=0.55\textwidth]{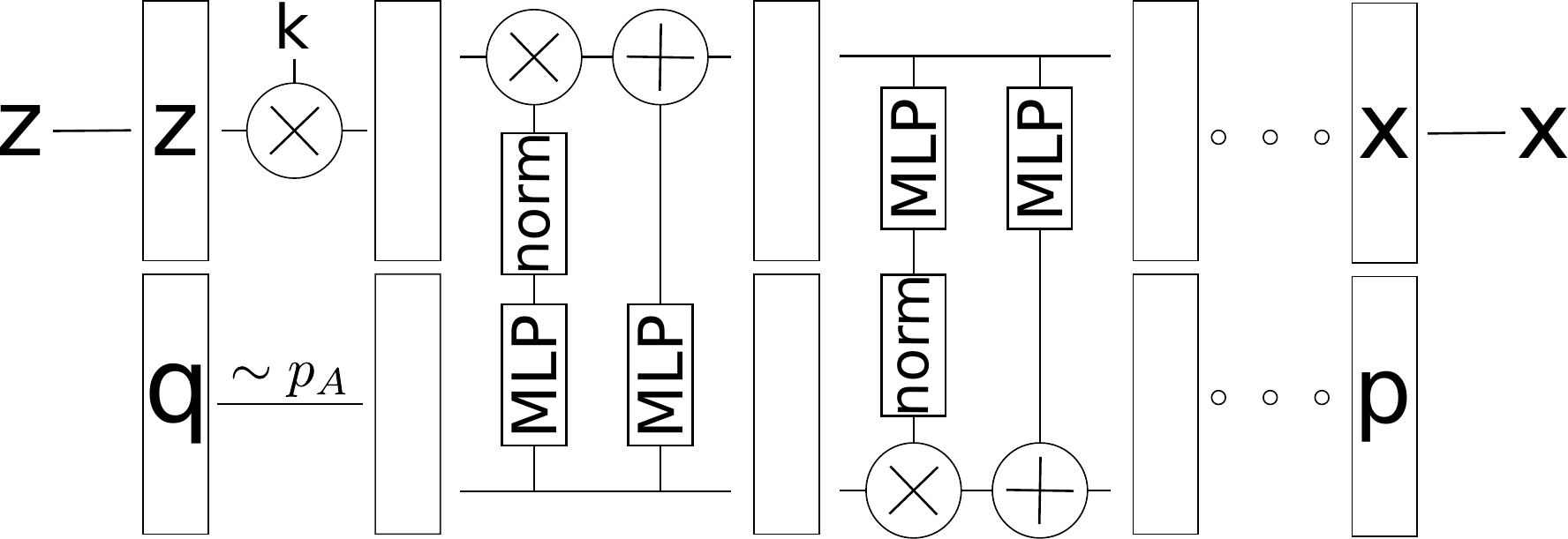} \caption{Schematic figure of the TSF architecture used here. Auxiliary momenta
$\vec{q}$ and coordinates $\vec{z}$ are coupled with volume preserving
networks where the outputs of the multi-layer perceptron (MLP) used
to generate the scaling variables are normalized. The first layer
multiplies the latent space coordinates $\vec{z}$ with a scalar factor
$k$, which adjusts for difference in entropy between latent and phase
space.}
\label{fig:hf_scheme}
\end{figure}

The recently proposed stochastic normalizing flows (SNF) \cite{wu2020stochastic}
can be used to construct TSFs. Here we utilize a type of SNF which
is motivated by Hamiltonian Monte Carlo (HMC) and utilizes operations
in an augmented space, where the auxiliary momenta are distributed
according to a distribution $p_{A}$. However, in contrast to the
original HMC method \cite{duane1987hybrid}, we do not propagate the
system by Hamiltonian dynamics, but learn a (deterministic) flow $D$
that takes this role. This flow shares similarities to other augmented
flows, such as Hamiltonian flows \cite{Greydanus2019HamiltonianNN,Toth2019HamiltonianGN}
and Augmented Normalizing Flows (ANFs) \cite{Huang2020AugmentedNF}.
To sample from the model, i.e. $\vec{x}=f_{\tau}(\vec{z});\vec{z}\sim p_{Z}^{\tau}(\vec{z})$,
the flow consists of three consecutive steps
\begin{enumerate}
\item Sample auxiliary momenta $\vec{q}\sim p_{A}^{\tau}(\vec{q})$, and
define the point in phase space $\vec{\upsilon}=(\vec{z},\vec{q})$
\item Propagate the point in phase space by the dynamics $\vec{\gamma}=(\vec{x},\vec{p})=D(\vec{\upsilon})$
\item Project onto the configuration variables $\vec{x}$
\end{enumerate}
With the flow described above we can construct a TSF by choosing $p_{A}^{\tau}(\vec{q})=\mathcal{N}(\vec{q}|0,\tau)$
and choosing a dynamics $D$ with a constant Jacobian which satisfies
the temperature scaling condition (Eq. \ref{eq:scaling_condition}).
A convenient way of constructing volume preserving dynamics, i.e.
$\left|\det J_{D}(\vec{\upsilon})\right|=1$, is obtained by altering
the RNVP network structure \cite{dinh2016density}, such that the
product of the outputs of the scaling layers is equal to unity. This
is done by subtracting the mean of the log outputs from each scaling
layer, similar to \cite{Sorrenson2020DisentanglementBN}. In addition
to the volume preserving RNVP layers we scale the latent space coordinates
by a trainable scalar, which allows to adjust for entropy difference
between the prior and the target. The resulting Jacobian factor is
a constant and, hence, still fulfills the scaling condition. The flow
architecture is shown in Fig. \ref{fig:hf_scheme}.

This architecture can also be viewed as an instance of an ANF, where
the augmented prior distribution $p_{\Upsilon}^{\tau}\left(\vec{\upsilon}\right)=p_{Z}^{\tau}(\vec{z})p_{A}^{\tau}(\vec{q})$
is mapped to the joint output distribution $p_{\Gamma}^{\tau}\left(\vec{\gamma}\right)$
via the invertible dynamics $D.$ The joint target distribution is
given by $\mu_{X}^{\tau}(\vec{x})p_{A}^{\tau}(\vec{p})$. As the flow
fulfills the temperature scaling condition, a temperature change of
the prior, i.e. $\tau\to\tau'$, will change the output accordingly.
In the case of a factorized output distribution $p_{\Gamma}^{\tau}\left(\vec{\gamma}\right)=p_{X}^{\tau}\left(\vec{x}\right)p_{A}^{\tau}\left(\vec{p}\right)$
the marginal distribution $p_{X}^{\tau}\left(\vec{x}\right)$ is scaled
correctly as well. This is ensured if the joint target distribution
is matched correctly.

\paragraph{Training}

As in \cite{noe2019boltzmann}, the TSFs are trained by a combination
of a maximum-likelihood (forward KL) and energy-based (reverse KL)
loss $\mathcal{L}=\left(1-\lambda\right)\mathcal{L}_{ML}+\lambda\mathcal{L}_{KL}$,
where the mixing parameter $\lambda$ is increased from zero during
training. The maximum likelihood loss minimizes the negative log likelihood
(nll), which requires samples from a distribution $p'_{\Gamma}(\gamma)$
that is similar to the target distribution
\begin{equation}
\mathcal{L}_{ML}=\mathbb{E}_{\vec{\gamma}\sim p'_{\Gamma}(\vec{\gamma})}\left[-\log p_{\Upsilon}(D^{-1}(\vec{\gamma}))-\log\left|\det J_{D^{-1}}(\vec{\gamma})\right|\right]\label{eq:nll}
\end{equation}
\textit{\emph{As the target energy $U(x)$}}\textit{ }\textit{\emph{is
defined by the physical system of interest, we can also rely on en}}ergy-based
training where the reverse KL divergence is minimized via the variational
free energy 
\begin{equation}
\mathcal{L}_{KL}=\mathbb{E}_{\vec{\upsilon}\sim p_{\Upsilon}(\vec{\upsilon})}\left[U(D(\vec{\upsilon}))-\log\left|\det J_{D}(\vec{\upsilon})\right|\right]
\end{equation}

\paragraph{Sampling and latent Monte Carlo}

One way to generate samples using the TSF is by directly sampling
from the Gaussian prior at the desired temperature and transforming
the samples to configuration space. As the TSF is only an approximation
to the Boltzmann distribution, these samples will generally be biased.

In order to ensure that the TSF generates unbiased samples from (\ref{eq:canonical_density}),
we use it in a MCMC framework. A proposal $\vec{x}'$ is generated
from configuration $\vec{x}$ by sampling auxiliary momenta $\vec{p}\sim p_{A}^{\tau}\left(\vec{p}\right)$
to define $\vec{\gamma}=(\vec{x},\vec{p})$, then applying the inverse
dynamics $\vec{\upsilon}=D^{-1}(\vec{\gamma})$ followed by a random
displacement $\vec{\upsilon}'=\vec{\upsilon}+\vec{\xi}$, with $\vec{\xi}\sim\mathcal{N}(0,\sigma^{2})$,
in latent space and then transforming back into configuration space
$(\vec{x}',\vec{p}')=D(\vec{\upsilon}')$. Accepting such a step with
acceptance probability $p_{\text{acc}}^{\tau}\left((\vec{x},\vec{p})\to(\vec{x}',\vec{p}')\right)=\min\left\{ 1,\exp\left[-\left(U(\vec{x}')-U(\vec{x})+\left\Vert \vec{p}'\right\Vert ^{2}/2-\left\Vert \vec{p}\right\Vert ^{2}/2\right)/\tau\right]\right\} $
enforces detailed balance in configuration space and thus ensures
convergence to the Boltzmann distribution.

As the TSF is able to generate distributions at several temperatures,
we can combine the MCMC moves with PT \cite{swendsen1986replica,Geyer1991MarkovCM,Hukushima1996ExchangeMC}.
In PT, sampling is performed at a set of temperatures in parallel.
Additionally to TSF-MCMC steps, samples can randomly be exchanged
between two randomly chosen temperatures. This allows the sampler
to overcome energy barriers quickly at high temperatures while still
preserving details of the wells at lower temperatures.

\section{Experiments}

We carry out experiments at two different test models. Firstly, we
demonstrate that the TSF drastically improves the temperature scaling
as compared to the original RNVP. In a second experiment we show that
for the Ala2 molecule the TSF is able to generate samples close to
the Boltzmann distribution and that the MCMC scheme results in unbiased
samples.

\paragraph{Mixture of multi-dimensional double wells}

\begin{figure}
\begin{centering}
\includegraphics[width=0.8\textwidth]{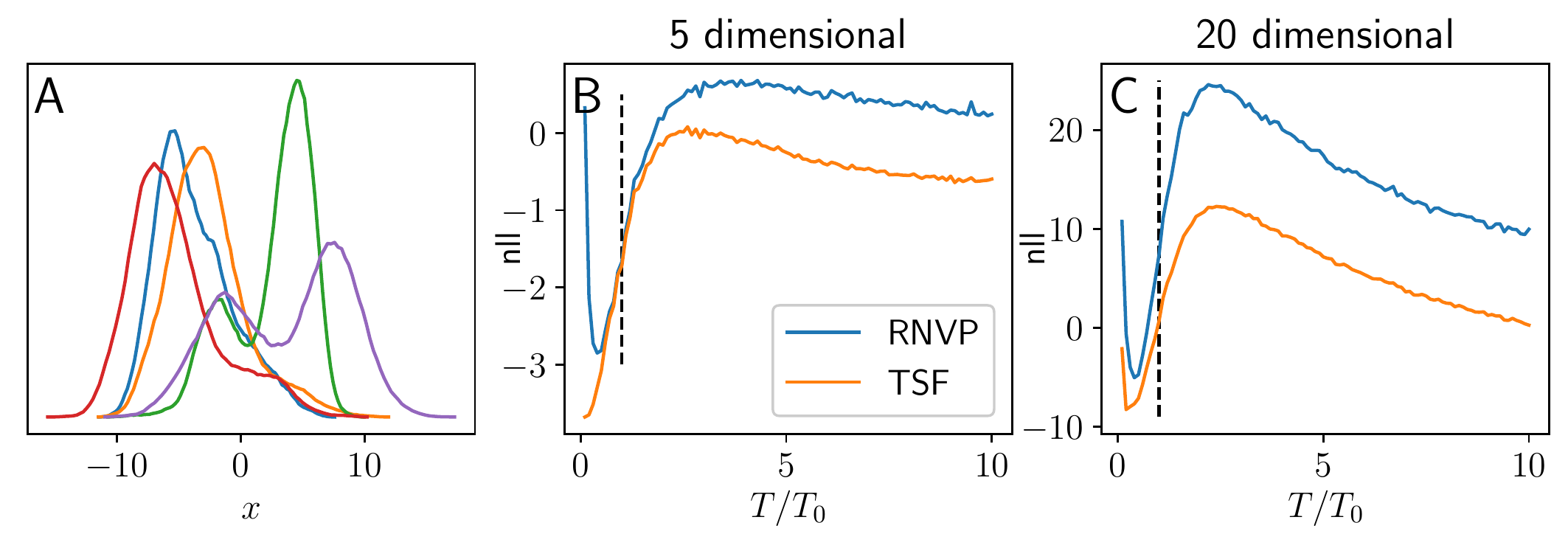}
\par\end{centering}
\caption{Results for the multi-dimensional double well system. \textbf{A}:
marginal density of the first 5 coordinates of the 20d system. \textbf{B:}
comparison of the nll as a function of the temperature between the
TSF and the RNVP network structure in the 5 dimensional case (lower
is better). The dashed line indicates the training temperature $T=1$
\textbf{C}: same as B but for the 20 dimensional system.}

\label{Fig.:multi-d-system}
\end{figure}
As a toy system that can easily be formulated in different dimensions,
we select a mixture of multi-dimensional double wells which are mixed
via a correlation matrix. The energy of this system is given by $U(x)=U_{\text{dw}}(A\vec{x}),$
with $U_{\text{dw}}(\vec{x})=\sum_{i}^{d}a_{i}x_{i}+bx_{i}^{2}+cx_{i}{}^{4}$.
The parameter vector $\vec{a}$ and correlation matrix $A$ are chosen
at random. We study the system at dimensions $d=\{5,20\}$ and compare
the TSF with a RNVP flow with comparable numbers of trainable parameters.
Fig. \ref{Fig.:multi-d-system} left shows the marginal density of
the first 5 coordinates of the 20d system. The Boltzmann Generators
are trained at $T=1$ and then analyzed by comparing the nll (Eq.
\ref{eq:nll}) of equilibrium samples, which were generated by Gaussian
increment MCMC in combination with PT at $100$ temperatures in the
range $T=0.1$ to $T=10$ (Fig. \ref{Fig.:multi-d-system} center
and right). For the 5d system we observe that both network structures
perform equally well in the close vicinity of the training temperature,
however the TSF has a significantly lower nll at temperatures further
away from the ones it was trained on. In the case of the 20d system,
the TSF consistently outperforms the RNVP even at the training temperature.
This indicates that the TSF is a more expressive network structure
with stronger temperature scaling.

\paragraph{Alanine Dipeptide}

\begin{figure}
\begin{centering}
\includegraphics[width=0.89\textwidth]{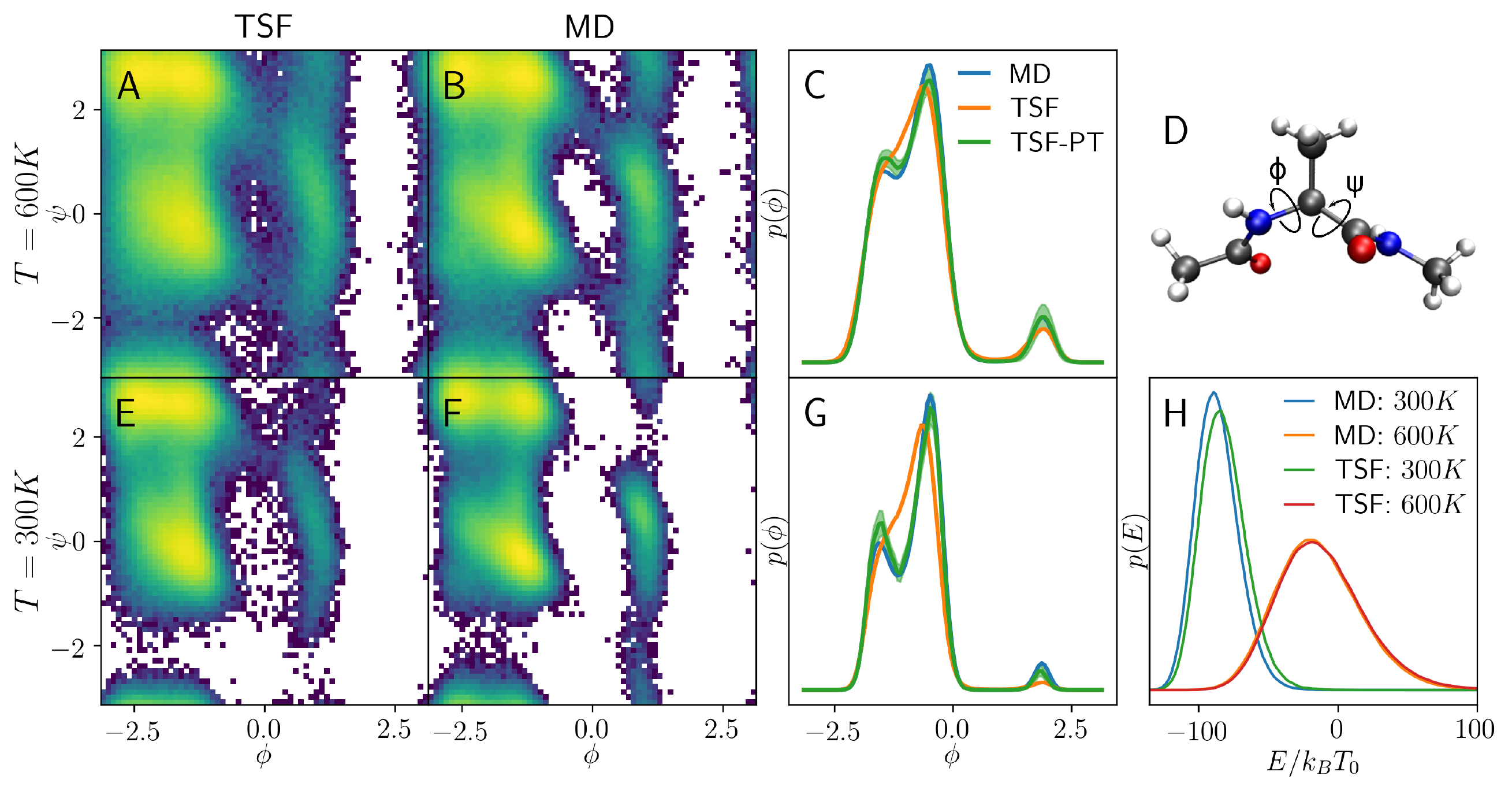}
\par\end{centering}
\label{fig:ala2}

\caption{Results for Alanine Dipeptide (\textbf{D}) in implicit solvent. \textbf{A,
B, E, F:} Density of $\phi,\psi$ variables from different methods.
\textbf{A: }TSF samples generated at the training temperature of $T=600$
K (ground truth is\textbf{ B}). \textbf{E:} Samples generated with
the same TSF at $T=300$ K to be compared (ground truth is \textbf{F}).
Histograms of the dihedral angle\textbf{ $\phi$ }are shown in \textbf{C
}for $T=600$ K and in \textbf{G }for $T=300$ K. \textbf{H:} Energy
histograms at the different temperatures. \textbf{D:} Alanine Dipeptide
molecule}
\end{figure}
We further test TSF on the Alanine Dipeptide molecule in an implicit
solvent model. For this system we use an invertible coordinate transformation
layer and operate the TSF on a representation of the molecule in terms
of distances and angles. As we are specifically interested in the
temperature scaling aspect, we generate sets of samples at temperatures
$T=\{300\:\mathrm{K},600\:\mathrm{K}\}$ using MD. Our goal is to
use the samples at $T=600\:\mathrm{K}$ to train the TSF and then
use the TSF to sample at $T=300\:\mathrm{K}$. The MD set at $T=300\:\mathrm{K}$
serves as ground truth for comparison. We use the TSF to generate
samples in configuration space and compare the Ramachandran plots
and distributions of the $\phi$ angle (Fig. \ref{fig:ala2}). We
observe good agreement at the training temperature (Fig. \ref{fig:ala2}
panels A to C). At $T=300\:\mathrm{K}$ (Fig. \ref{fig:ala2} panels
E to G) the TSF still finds the major minima at around $\phi\approx-2$,
but under-samples the minimum at $\phi\approx1$. This deviation from
the target distribution is likely stemming from limited expressivity
of the flow. We are able to recover the correct distribution of $\phi$
when using the Monte Carlo scheme in a PT fashion (Fig. \ref{fig:ala2}
C and G). Furthermore, the output of the TSF closely recovers the
distribution of energies (Fig. \ref{fig:ala2} H) at both temperatures.

\section{Discussion}

In this work, we derived and constructed temperature-steerable flows
(TSF) that correctly scale the output distribution of a BG with temperature.
We showed that this novel type of flow can be used to train a BG at
one temperature and generate distributions at other temperatures.
Further progress could be made by combining samples at different temperatures
when collecting training data and thus improve the quality of the
BG. Another application could be the investigation of temperature
dependent observables, e.g. magnetization in spin systems or free
energy difference in proteins.

\section{Broader impact}

Sampling equilibrium samples is important for physics and other fields.
Applications range from development of new materials to drug discovery.
A lot of computational resources and energy is required for these
simulations. Enhancing such sampling algorithms with Machine Learning
methods has the potential to increase the efficiency. Allowing to
reduce the computational cost and thus energy consumption significantly.
One promising approach to tackle this problem are the framework of
Boltzmann Generators (BGs), to which this paper contributes. This
paper enables BGs to generate Boltzmann Distributions at different
temperatures. This allows them to be trained at high temperatures
where sampling is comparably cheap and furthermore to be used in enhanced
sampling algorithms such as parallel tempering.

One risk of this method is that enabling and improving sampling methods
for which currently no convergence criterion is known can lead to
false claims. In the context of flow bases samplers ergodicity is
not well studied. While broken ergodicity can also be an issue in
MD and MCMC, these are known to be ergodic at least in the asymptotic
limit, i.e. the limit of infinitely long simulations. This needs to
be further researched to arrive at criteria that guarantee convergence.

\section{Acknowledgments}

We gratefully acknowledge support by the Deutsche Forschungsgemeinschaft
(SFB1114, project C03), the European research council (ERC CoG 772230
\textquotedblleft ScaleCell\textquotedblright ), the Berlin Mathematics
center MATH+ (AA1-6), and the German Ministry for Education and Research
(BIFOLD - Berlin Institute for the Foundations of Learning and Data).
We thank Jonas Köhler, Andreas Krämer and Yaoyi Chen for insightful
discussions.

\bibliographystyle{plain}
\bibliography{literature}
\textit{}
\end{document}